\documentclass[runningheads]{llncs}

\usepackage[T1]{fontenc}
\usepackage{graphicx}
\usepackage{amsmath,amssymb}
\usepackage{booktabs}
\usepackage{multirow}
\usepackage{cite}
\usepackage{color} 
\usepackage{subfigure}
\usepackage[hidelinks]{hyperref} 

\graphicspath{{fig/}}
\begin{document}

% ==========================================================
% 1. 论文标题与作者信息
% ==========================================================
\title{Bridging Semantics and Signals: An LLM-Driven MoE Framework for Radio Map Prediction}

\titlerunning{LLM-MoE for Radio Map Prediction}

\author{Qi Yang\inst{1,2} \and
Xinyue Jiang\inst{1,2} \and
Tong Li\inst{1,2} \and \\
Chanchan Zhu* \inst{3} \and 
Shaolong Gu \inst{3} \and 
Zhu Xiao\inst{1,2}
}
\authorrunning{Q. Yang et al.}
\institute{College of Computer Science and Electronic Engineering, Hunan University. \and
Ministry of Education Key Laboratory of ``Fusion Computing of Supercomputing and Artificial Intelligence'', Changsha, China. \and The 54th Research Institute of China Electronics Technology Group Corporation (CETC),
Shijiazhuang, Hebei 050081, China.\footnotetext{* Corresponding author}\\
\email{tliay@hnu.edu.cn, chanchannibang1988@126.com, Zhxiao@hnu.edu.cn}\\
}

\maketitle
\begin{abstract}
High-fidelity radio map prediction is essential for wireless network optimization, yet existing single-network models frequently fail to generalize across diverse multi-band and multi-terrain scenarios due to rigid parameter constraints. We introduce LLM-MoE, a framework that employs a divide-and-conquer Mixture-of-Experts strategy to dynamically allocate modeling tasks to matched expert sub-networks. Our approach features a heterogeneous expert pool for characterizing varied physical propagation behaviors, combined with an LLM-driven semantic extractor that provides global physical awareness. A cross-modal fusion module integrates these macroscopic semantics with local spatial features for adaptive routing. Experimental results on the SpectrumNet dataset show that LLM-MoE outperforms top-tier baselines, achieving a 10.0\% reduction in Mean Squared Error (MSE) and 9.7\% in Normalized Mean Squared Error (NMSE), while ensuring robust generalization across heterogeneous environments.

\end{abstract}

\section{Introduction}
\label{sec:intro}

Radio maps characterize the spatial distribution of electromagnetic features, such as signal strength and path loss, across a target area~\cite{ref_jiang_unirm, CSI_ICC}. By providing fine-grained insights into multipath attenuation and interference hotspots, high-precision radio map prediction becomes a vital prerequisite for technologies such as intelligent network planning and dynamic spectrum optimization~\cite{PEFNet, liu2024coverage, huang2023safe}. 
Data-driven deep learning has become the primary paradigm for radio map prediction. Different neural architectures are uniquely suited to capture specific propagation mechanisms: Convolutional Neural Networks (CNNs)~\cite{su2025jointly, liu2024digital} excel at extracting local diffraction patterns, Vision Transformers (ViTs) model long-range spatial dependencies, and Multilayer Perceptrons (MLPs)~\cite{su2025gennet} effectively fit distance-based non-linear path loss. However, real-world wireless environments are highly heterogeneous, spanning diverse scenarios such as urban multipath fading, suburban scattering, and rural line-of-sight propagation~\cite{huang2026l2s, liu2026joint, li2023carbon}. Bound by fixed structural biases, single-architecture models struggle to adapt to these vastly different mechanisms simultaneously, creating a capacity bottleneck that restricts their generalization across varied scenarios.

To address these limitations, we propose a Large Language Model (LLM)-driven Mixture-of-Experts (MoE) framework for radio map prediction. The MoE framework employs a dynamic routing mechanism to allocate inputs to specialized subnetworks, serving as an efficient divide-and-conquer solution. We construct a heterogeneous expert pool consisting of CNN, ViT, MLP, and Graph Neural Network (GNN) architectures to adaptively model diverse propagation mechanisms. This approach ensures that each modeling task is assigned to the most suitable expert, thereby reducing the fitting biases associated with monolithic, single-network models. Additionally, we utilize an LLM to extract macroscopic semantic information by converting environmental attributes into structured textual prompts. These semantics are integrated with local spatial features through a specialized cross-modal fusion module, enabling a semantics-guided dynamic routing mechanism for the heterogeneous expert pool. This LLM-driven fusion enables the gating network to adaptively allocate inputs to the most appropriate experts, bridging the gap between localized spatial signals and the global environmental context and enhancing cross-scenario generalization.

The main contributions of this paper are summarized as follows:

$\bullet$ We investigate radio map prediction across diverse wireless environments. To the best of our knowledge, we are the first to introduce an LLM-driven MoE framework into radio map prediction, utilizing a divide-and-conquer strategy to decouple the modeling of various physical mechanisms, thereby mitigating the capacity limits of single networks.
    
$\bullet$ We propose the LLM-MoE framework, featuring three core components: a heterogeneous expert pool to adaptively model diverse propagation mechanisms; an LLM-driven module to extract macroscopic semantic descriptions; and a cross-modal gating network for semantics-guided dynamic routing. This design effectively addresses the capacity bottlenecks of single networks and the semantic deficiency issues of traditional MoE architectures.

$\bullet$ We conduct extensive experiments on the open-source SpectrumNet dataset to evaluate LLM-MoE's performance. The results demonstrate that LLM-MoE outperforms state-of-the-art baselines, reducing the Mean Squared Error (MSE) by 10.0\% and the Normalized Mean Squared Error (NMSE) by 9.7\% under strictly aligned five-band data.

\section{Related Work}
\label{sec:related_work}

Data-driven approaches have become the dominant paradigm for radio map prediction due to their strong ability to model complex nonlinear relationships and enable efficient inference~\cite{Zhou20263DCKM, bai2026passive}. These methods leverage neural networks to learn propagation patterns and signal strength distributions directly from data, without relying on explicit statistical assumptions~\cite{chai2024generative, li2025generative}. For instance, Levie \textit{et al.}~\cite{ref_levie_radiounet} proposed RadioUNet, a CNN-based model that was among the first to approximate path loss across diverse environments. This line of work has been further extended by a series of UNet-based architectures~\cite{PMNet, PEFNet, ref_jiang_unirm}, which enhance multi-scale feature extraction for improved radio map prediction accuracy. Beyond CNNs, alternative neural architectures have also been explored. Autoencoders~\cite{ref_ae} have been used to exploit the underlying manifold structure of radio maps, while generative models~\cite{ref_phyrmdm} aim to capture the underlying distribution of radio maps. Transformer-based models~\cite{liu2025paying} have also been introduced to model long-range spatial dependencies, such as Loc-SwinUNet. In addition, GNNs~\cite{ref_radiogat} have been investigated to capture relational structures in wireless environments. Despite the success of these diverse architectures, most existing methods rely on a single neural network with fixed structural biases. Such models are inherently limited in their ability to simultaneously capture heterogeneous electromagnetic propagation behaviors across different frequency bands and terrain scenarios. Their performance often degrades under cross-scenario or cross-frequency conditions, underscoring the need for more flexible, adaptive modeling frameworks.

\section{System Model}
\label{sec:system_model}

Consider a target geographical region partitioned into a discrete $W \times H$ grid, where each cell corresponds to a specific spatial coordinate. The physical layout of this region is characterized by a topology matrix $\mathbf{X}_{topo}$ representing building layouts, and a position matrix $\mathbf{X}_{tx}$ representing the locations of $N_t$ radio transmitters. In practice, the system can only obtain a limited set of $K$ measurement samples. These sparse observations are denoted by a measurement matrix $\mathbf{X}_{sparse} \in \mathbb{R}^{W \times H}$. To explicitly quantify spatial geometric attenuation, a physical depth map $\mathbf{X}_{depth} \in \mathbb{R}^{W \times H}$ is further introduced. These spatial modalities are concatenated along the channel dimension to form a visual input tensor $\mathbf{X}_{img} = [\mathbf{X}_{topo}, \mathbf{X}_{tx}, \mathbf{X}_{depth}, \mathbf{X}_{sparse}]$. The core objective is to design a neural network $\mathcal{F}_{\Theta}$, parameterized by $\Theta$, to establish a robust mapping from these multi-modal observations, i.e., physical layout, transmitter locations, depth map, and measurement samples, to the complete signal distribution. The model optimization aims to minimize the MSE between the predicted map $\hat{\mathbf{Y}}_f = \mathcal{F}_{\Theta}(\mathbf{X}_{img}, f)$ and the ground-truth map $\mathbf{Y}_f$, where $f$ denotes the operating frequency. The problem is mathematically formulated as:
\begin{equation}
\label{eq:problem_formulation}
\min_{\Theta} \quad \frac{1}{W \times H} \sum_{i=1}^{W} \sum_{j=1}^{H} \left( \mathbf{Y}_f(i,j) - \hat{\mathbf{Y}}_f(i,j) \right)^2,
\end{equation}
where $\mathbf{Y}_f(i,j)$ and $\hat{\mathbf{Y}}_f(i,j)$ represent the actual and predicted received signal power values at the grid coordinate $(i,j)$.
\section{Methodology}
\label{sec:model_design}

\subsection{Framework Overview}
\label{subsec:framework_overview}

\begin{figure}[tb]
    \centering
    \includegraphics[width=0.96\textwidth, height=0.4\textheight, keepaspectratio]{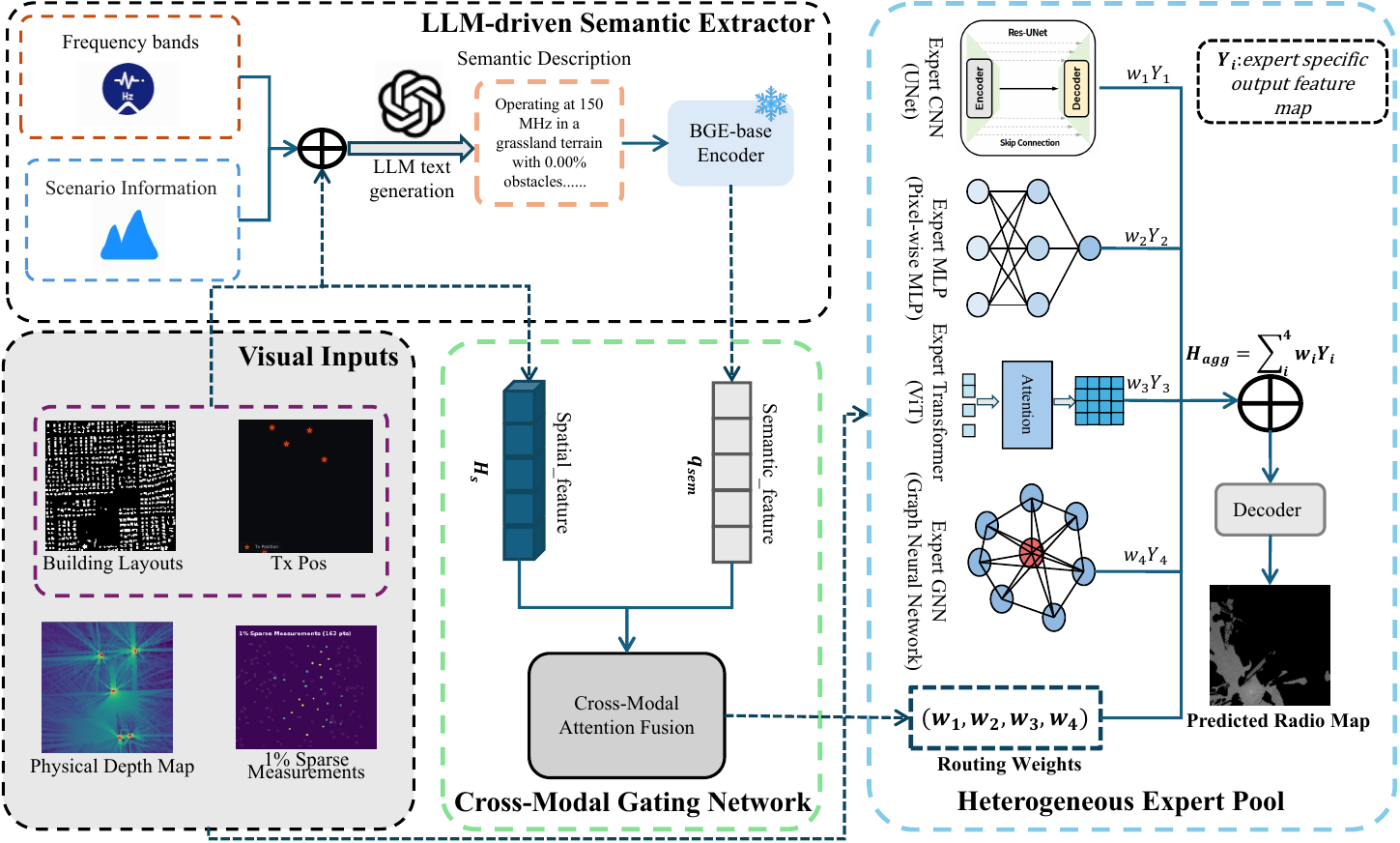}
    \caption{The overview of the LLM-MoE architecture, which consists of three core components: (i) Heterogeneous Expert Pool, (ii) LLM-driven Semantic Extractor, and (iii) Cross-Modal Gating Network.}
    \vspace{-2mm}
    \label{fig:framework}
\end{figure}

To effectively model electromagnetic propagation across diverse scenarios, LLM-MoE decouples the complex prediction task into specialized subtasks handled by dedicated expert networks. As illustrated in Fig.~\ref{fig:framework}, the framework consists of three core components: a Heterogeneous Expert Pool, an LLM-driven Semantic Extractor, and a Cross-Modal Gating Network.

\textbf{Heterogeneous Expert Pool}: This component comprises diverse neural network architectures designed to leverage distinct structural advantages. Different experts independently model specific physical propagation behaviors—such as local diffraction, scattering, and non-linear distance-based attenuation—allowing for more precise characterization than monolithic models.

\textbf{LLM-driven Semantic Extractor}: This module processes environmental attributes, including the number of transmitters, operating frequency bands, terrain categories, and overall building coverage ratios, alongside the building topology matrix. We utilize a pre-trained LLM for inference to extract a global semantic description, providing a macroscopic perspective that localized spatial features often lack.

\textbf{Cross-Modal Gating Network}: This network serves as the system's routing mechanism, integrating the extracted global semantic description with local spatial features to dynamically compute routing weights for each expert. This design ensures that routing decisions are informed by both localized topologies and macroscopic semantics.

\subsection{Heterogeneous Expert Pool}
\label{subsec:expert_pool}
Based on the MoE methodology, we design a heterogeneous expert pool to decouple complex electromagnetic propagation mechanisms. The core principle is to expand model capacity by routing inputs to specialized sub-networks, thereby handling highly differentiated scenarios without significantly increasing computational overhead. Our expert pool consists of four distinct neural network architectures. Specifically, given the unified visual input tensor $\mathbf{X}_{img}$, each $i$-th expert independently processes it to generate an expert-specific high-dimensional output feature map $\mathbf{Y}_i$. These output representations encapsulate distinct physical propagation patterns and serve as the foundational features for the subsequent weighted aggregation.

\textbf{CNN} is employed to capture local electromagnetic diffraction and short-range reflections. By utilizing a U-Net architecture with residual skip connections, it extracts hierarchical spatial features through successive downsampling and upsampling convolutions, making its structural bias highly effective at perceiving localized building occlusion patterns.

\textbf{Transformer} is utilized to model long-range spatial dependencies across complex building clusters. By implementing a ViT architecture that linearly projects input features into a sequence of non-overlapping patches, it employs a multi-head self-attention mechanism to capture global interactions, allowing the model to understand macroscopic topological relationships that exceed the limited receptive fields of standard convolutions.

\textbf{MLP} is employed to specifically model distance-based non-linear attenuation. By operating as a pixel-wise MLP, it decouples interference from the neighborhood receptive field, allowing the model to focus strictly on learning distance-dependent attenuation.

\textbf{GNN} is incorporated to simulate local electromagnetic scattering mechanisms by reconstructing spatial features into a graph $\mathcal{G} = (\mathcal{V}, \mathcal{E})$, where nodes represent spatial grids with Euclidean coordinates, and edges connect nodes within a valid physical scattering distance.

\subsection{LLM-Driven Semantic Extractor}
\label{subsec:semantic_extractor}

Relying solely on localized visual patches restricts the system's ability to perceive the macroscopic physical environment. Since radio wave propagation is heavily influenced by global factors such as the operating frequency band and surrounding terrain, we design an LLM-driven module to extract a comprehensive semantic description to provide the gating network with a global perspective.

We aggregate macroscopic attributes including the operating frequency $f$, the terrain category $c_{terrain}$, the number of transmitters and their position matrix $\mathbf{X}_{tx}$, the building topology matrix $\mathbf{X}_{topo}$, and the overall building coverage ratio, structuring them into natural language prompts. A pre-trained LLM then processes these prompts to generate a descriptive text $T_{sem}$, which goes beyond merely listing numerical parameters by encapsulating high-level physical deductions about the current scenario, such as the expected severity of multipath scattering or the dominant attenuation mechanisms, thereby effectively translating raw environmental parameters into a physical semantic description.

To map this unstructured text into a machine-readable format for the subsequent routing mechanism, we utilize the BGE-base encoder~\cite{ref_bge} to extract the high-dimensional semantic representation of the input text. Finally, to align this textual embedding with the visual latent space, we apply a projection network to compress it into the required $d$-dimensional final semantic feature vector $\mathbf{q}_{sem} \in \mathbb{R}^{1 \times d}$, effectively completing the cross-modal feature extraction without introducing unnecessary computational complexity.

\subsection{Cross-Modal Gating Network}
\label{subsec:gating_network}

\begin{figure}[tb]
    \centering
    \includegraphics[width=0.92\textwidth]{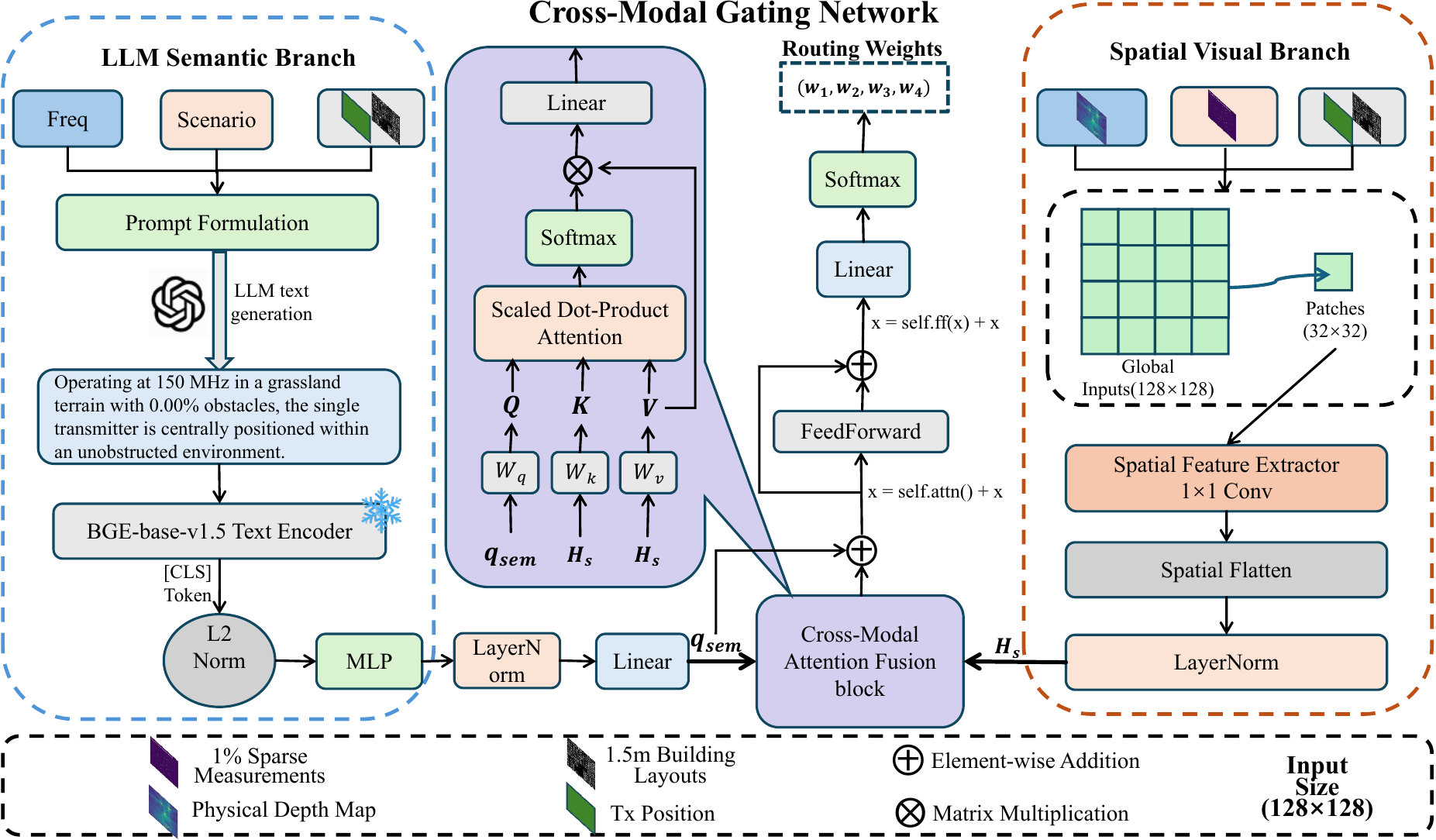} 
    \vspace{-3mm}
    \caption{Detailed architecture of the Cross-Modal Gating Network.}
    \vspace{-3mm}
    \label{fig:gating_network}
    \vspace{-3mm}
\end{figure}

Gating mechanisms in traditional MoE architectures primarily rely on linear projections of local spatial features, causing suboptimal routing decisions that lack global physical awareness. To overcome this limitation, we design a Cross-Modal Gating Network to facilitate semantics-guided dynamic expert allocation.

\textbf{Visual Feature Embedding.} As depicted in Fig.~\ref{fig:gating_network}, to align with the semantic feature dimensions, the visual input tensor $\mathbf{X}_{img}$ is sequentially processed by batch normalization and a $1 \times 1$ convolution for channel projection, and is subsequently flattened into a 1D sequence and subjected to layer normalization. Thus, we obtain the visual feature embedding $\mathbf{H}_s$. Notably, the visual input tensor $\mathbf{X}_{img}$ incorporates a physical depth map derived from the Euclidean distances and building occlusion states relative to the transmitters. This explicit spatial representation effectively assists the network in perceiving distance-dependent propagation characteristics.

\textbf{Cross-Modal Fusion Module.} To effectively integrate the global semantic description with local spatial features, we design a cross-modal fusion module where the final semantic feature vector $\mathbf{q}_{sem}$ is linearly mapped to the Query ($\mathbf{Q}$), while the spatial sequence feature $\mathbf{H}_s$ is linearly mapped to Keys ($\mathbf{K}$) and Values ($\mathbf{V}$):
\begin{equation}
\label{eq:cross_qkv}
\mathbf{Q} = \mathbf{q}_{sem}\mathbf{W}_Q, \quad \mathbf{K} = \mathbf{H}_s\mathbf{W}_K, \quad \mathbf{V} = \mathbf{H}_s\mathbf{W}_V,
\end{equation}
where $\mathbf{W}_Q$, $\mathbf{W}_K$, and $\mathbf{W}_V$ represent the learnable weight matrices for the linear projections. The cross-modal feature fusion mechanism is formulated as:
\begin{equation}
\label{eq:cross_attention}
\mathbf{H}_{fused} = \operatorname{softmax}\left(\frac{\mathbf{Q}\mathbf{K}^T}{\sqrt{d_k}}\right)\mathbf{V} \in \mathbb{R}^{1 \times d},
\end{equation}
where $d_k$ is the scaling factor matching the key dimension. This operation computes the correlation matrix between the global semantic description and all physical grid points, adaptively aggregating the extensive spatial features into a single scenario-aware fused representation $\mathbf{H}_{fused}$.

\textbf{Dynamic Routing and Decoder Aggregation.} The fused representation $\mathbf{H}_{fused}$ is then fed into a linear router that applies a Softmax function to output the routing weights $\mathbf{w} = [w_1, w_2, w_3, w_4]$ allocated to the four heterogeneous experts:
\begin{equation}
\label{eq:routing_weights}
\mathbf{w} = \operatorname{softmax}\left( \mathbf{H}_{fused}\mathbf{W}_{route} \right),
\end{equation}
where $\mathbf{W}_{route}$ is the learnable parameter matrix of the linear router.
Guided by this probability distribution, the system performs a dynamic weighted aggregation of the high-dimensional latent representations $\mathbf{Y}_i$ generated by the $i$-th expert, where each routing weight satisfies $w_i \in (0, 1)$ and $\sum_{i=1}^{4} w_i = 1$:
\begin{equation}
\label{eq:feature_aggregation}
\mathbf{H}_{agg} = \sum_{i=1}^{4} w_i \mathbf{Y}_i.
\end{equation}
Finally, the aggregated feature tensor $\mathbf{H}_{agg}$ is fed into a decoder comprising four residual blocks and a terminal convolutional layer, which progressively reduces the channel dimension while strictly preserving the spatial resolution to output the final predicted high-resolution radio map $\hat{\mathbf{Y}}_f \in \mathbb{R}^{W \times H}$.

\section{Experiments}
\label{sec:experiments}

\subsection{Experimental Settings}
\label{subsec:exp_settings}

\textbf{Dataset.} We evaluate the proposed LLM-MoE framework on the large-scale open-source SpectrumNet dataset~\cite{ref_spectrumnet}. Table~\ref{tab:dataset_properties} summarizes the detailed properties of the dataset. The original dataset provides radio maps at three altitudes. We extract a strictly aligned data subset covering five standard frequency bands and 11 terrain scenarios. In this dataset, the received signal power values are linearly normalized to $[0, 1]$ to facilitate stable training. We perform all model optimizations and metric calculations strictly within this dimensionless space.

\begin{table}[tb]
    \centering
    \scriptsize
    \caption{Properties of the SpectrumNet Dataset.}
    \vspace{-2mm}
    \label{tab:dataset_properties}
    \setlength{\tabcolsep}{3pt}
    \begin{tabular}{ll}
        \toprule
        \textbf{Properties} & \textbf{SpectrumNet} \\
        \midrule
        Real world maps & 15300 \\
        Projection area of each map & $1.28\text{ km} \times 1.28\text{ km}$ \\
        Spatial resolution & $10\text{ m}$ \\
        Scenario types & 11 \\
        Sample heights & 1.5 m, 30 m, 200 m \\
        Frequency bands & 150 MHz, 1.5 GHz, 1.7 GHz, 3.5 GHz, 22 GHz \\
        \bottomrule
    \end{tabular}
    \vspace{-2mm}
\end{table}

\textbf{Baselines and Evaluation Metrics.} We compare the proposed LLM-MoE with state-of-the-art models to evaluate its performance and generalization. The selected baselines include {RadioUNet}~\cite{ref_levie_radiounet}, {PMNet}~\cite{PMNet}, {PEFNet}~\cite{PEFNet}, {AE}~\cite{ref_ae}, and 
{PhyRMDM}~\cite{ref_phyrmdm}. To evaluate the reconstruction quality, we employ four standard metrics: MSE, Root Mean Squared Error (RMSE), NMSE, and Peak Signal-to-Noise Ratio (PSNR).

\subsection{Performance Evaluation}

\begin{table}[tb]
  \centering
  \scriptsize
  \caption{Overall Performance Comparison with Baselines}
  \vspace{-3mm}
  \label{tab:overall_performance}
  \setlength{\tabcolsep}{6pt}
  \begin{tabular}{lcccc}
    \toprule
    \textbf{Methods} & \textbf{MSE} $\downarrow$ & \textbf{NMSE} $\downarrow$ & \textbf{RMSE} $\downarrow$ & \textbf{PSNR} $\uparrow$ \\
    \midrule
    AE        & 0.02211 & 0.31320 & 0.13967 & 17.75 \\
    RadioUNet & 0.02029 & 0.27210 & 0.13333 & 18.17 \\
    PMNet     & 0.02101 & 0.28676 & 0.13670 & 17.88 \\
    PEFNet    & \underline{0.02020} & \underline{0.26934} & \underline{0.13297} & \underline{18.20} \\
    PhyRMDM   & 0.02032 & 0.27319 & 0.13309 & 18.19 \\
    \textbf{LLM-MoE (Ours)} & \textbf{0.01819} & \textbf{0.24313} & \textbf{0.12183} & \textbf{19.72} \\
    \midrule
    \textbf{Improv.} & \textbf{10.0\%} & \textbf{9.7\%} & \textbf{8.4\%} & \textbf{8.4\%} \\
    \bottomrule
  \end{tabular}
  \vspace{-6mm}
\end{table}

\textbf{Overall Performance.} Table~\ref{tab:overall_performance} summarizes the test results across all frequency bands and scenarios. LLM-MoE consistently outperforms all evaluated baselines on all metrics. It achieves the lowest MSE of 0.01819, NMSE of 0.24313, and RMSE of 0.12183 while obtaining the highest PSNR of 19.72 dB, demonstrating clear overall superiority. Compared with the second-best baseline PEFNet, LLM-MoE reduces the MSE by 10.0\% and improves the PSNR by 8.4\%. These results indicate that the proposed method not only reduces prediction errors but also enhances the structural fidelity of the reconstructed radio maps.

\begin{figure}[tb]
  \centering
  \vspace{-2mm}
  % 第一行：Ocean
  \subfigure[{\scriptsize Ocean (150 MHz)}]{\includegraphics[width=0.3\textwidth]{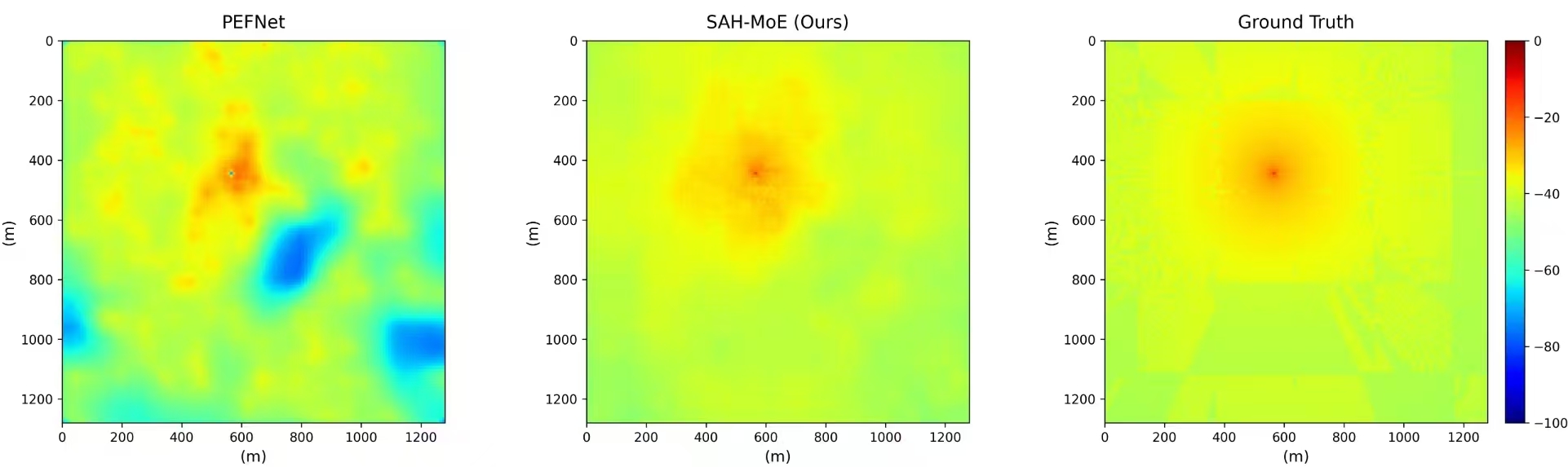}}
  \hfill
  \subfigure[{\scriptsize Ocean (1.7 GHz)}]{\includegraphics[width=0.3\textwidth]{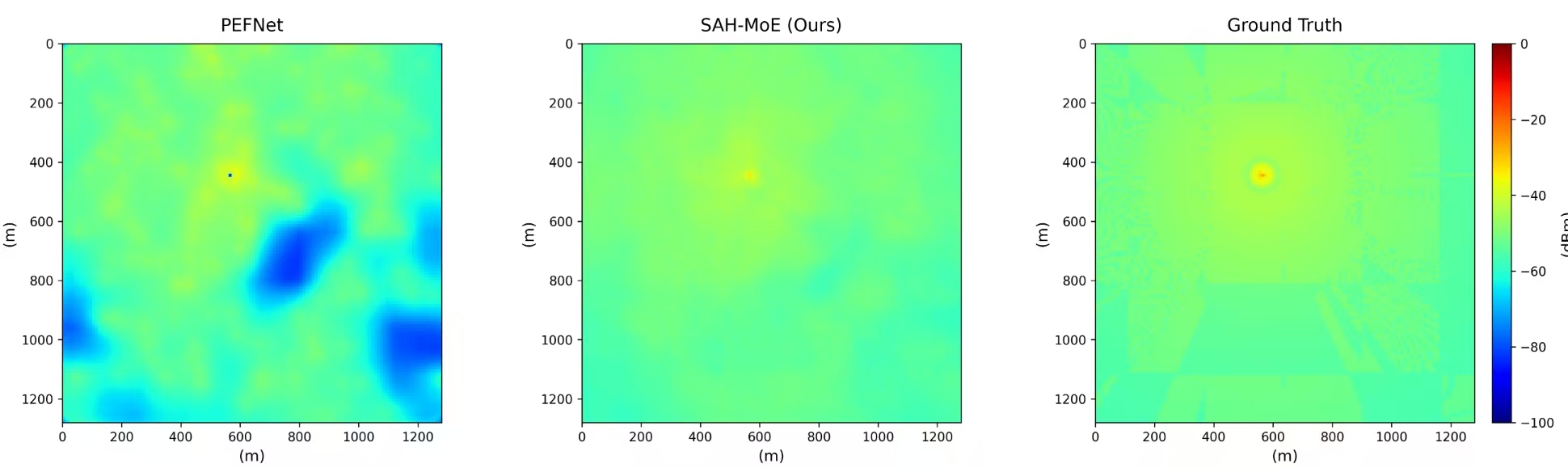}}
  \hfill
  \subfigure[{\scriptsize Ocean (22 GHz)}]{\includegraphics[width=0.3\textwidth]{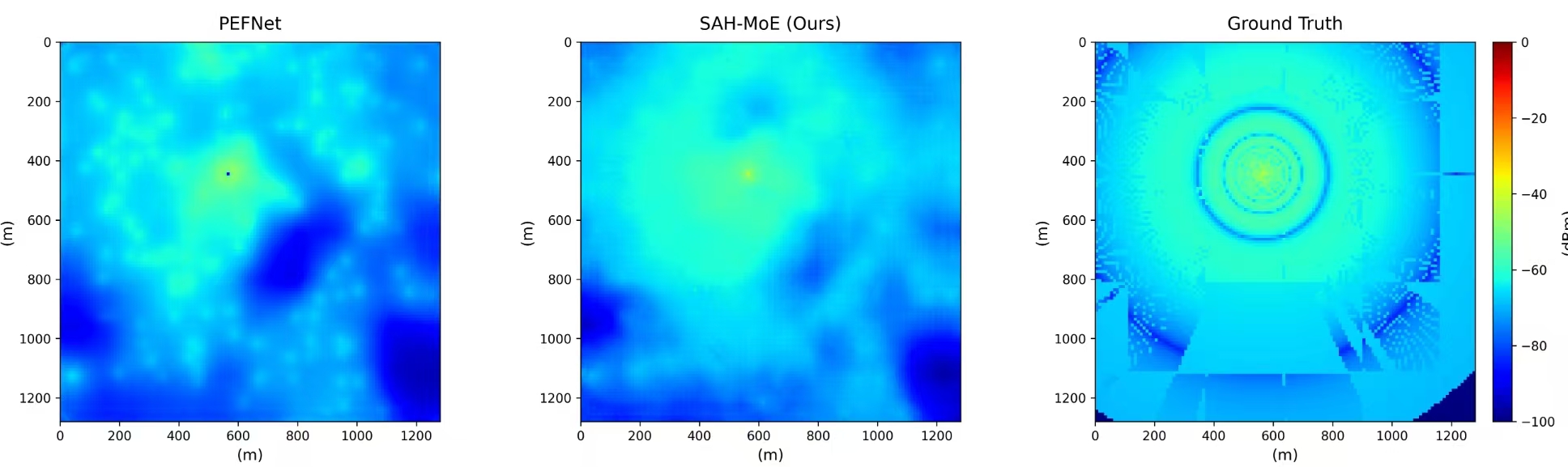}}
  \\ 
  \vspace{-2mm}   
  % 第二行：Island
  \subfigure[{\scriptsize Island (150 MHz)}]{\includegraphics[width=0.3\textwidth]{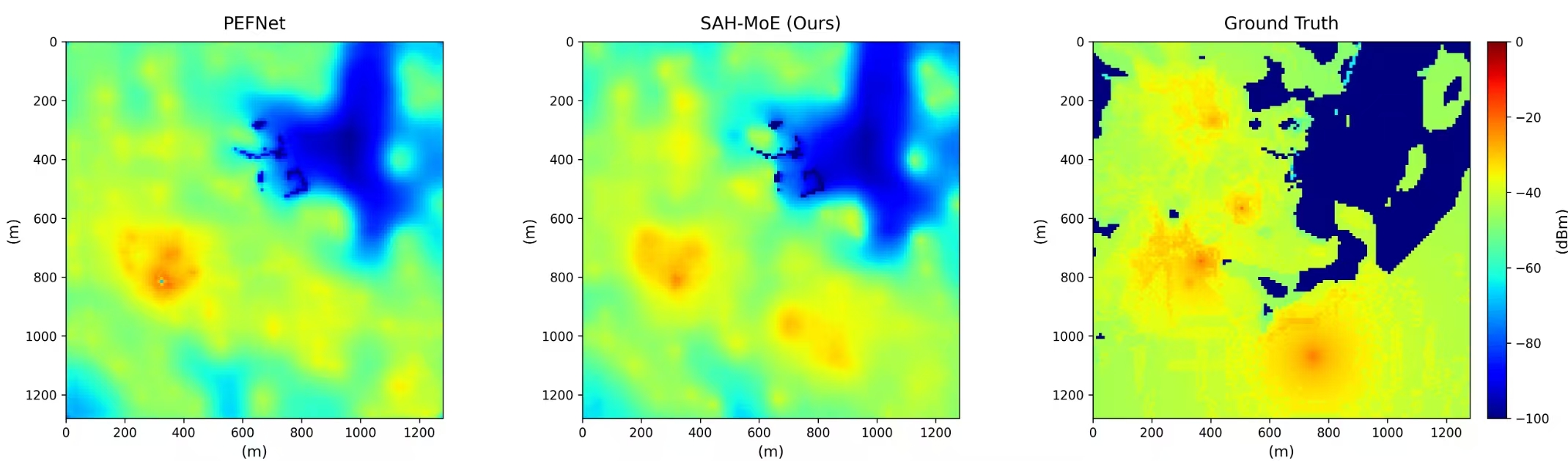}}
  \hfill
  \subfigure[{\scriptsize Island (1.7 GHz)}]{\includegraphics[width=0.3\textwidth]{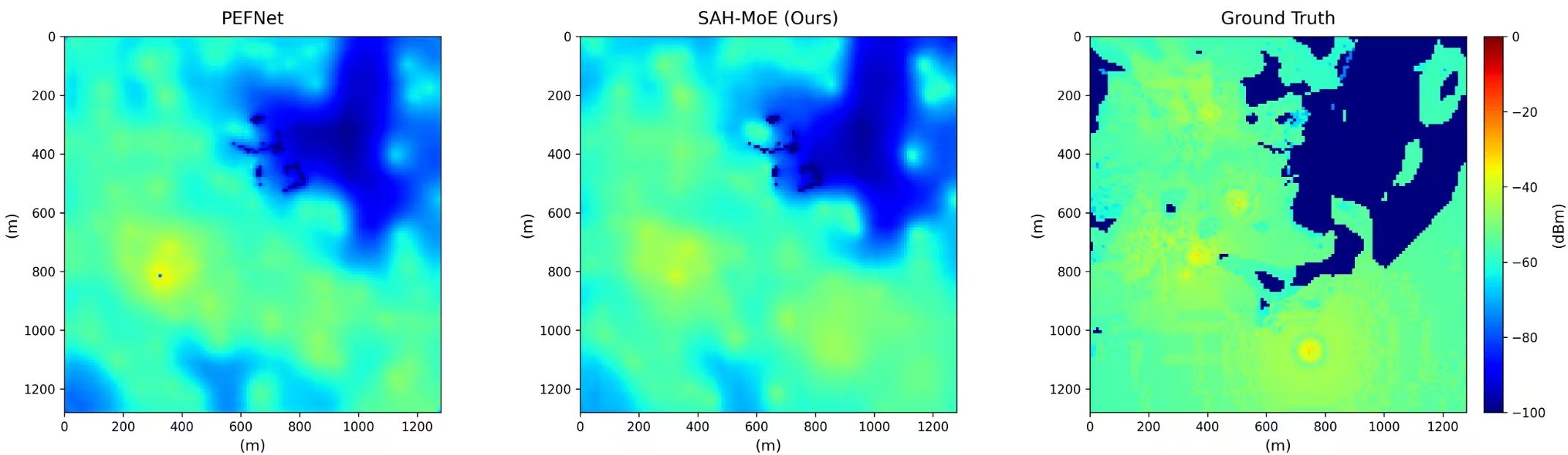}}
  \hfill
  \subfigure[{\scriptsize Island (22 GHz)}]{\includegraphics[width=0.3\textwidth]{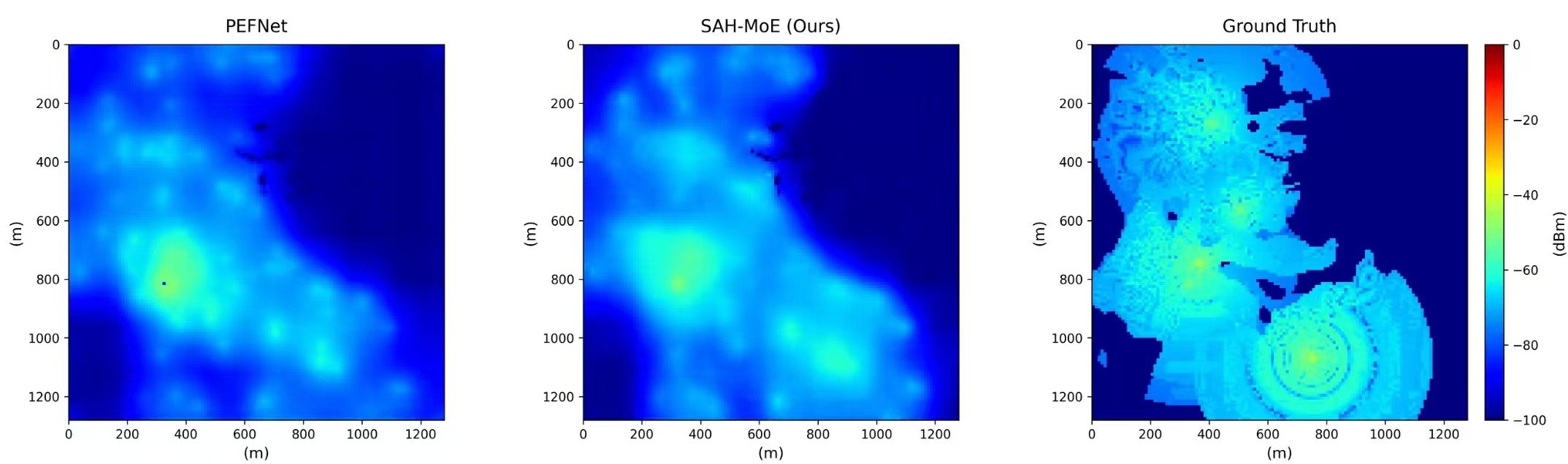}}
  \\
  \vspace{-2mm} 
  
  % 第三行：Dense Urban
  \subfigure[{\scriptsize Dense Urban (150 MHz)}]{\includegraphics[width=0.3\textwidth]{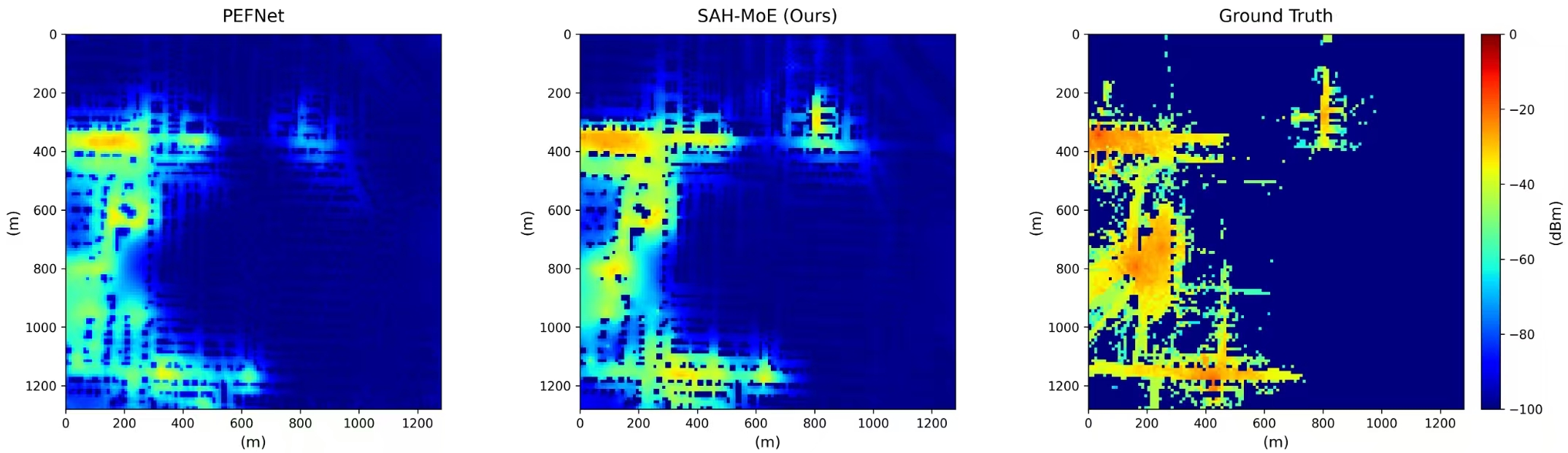}}
  \hfill
  \subfigure[{\scriptsize Dense Urban (1.7 GHz)}]{\includegraphics[width=0.3\textwidth]{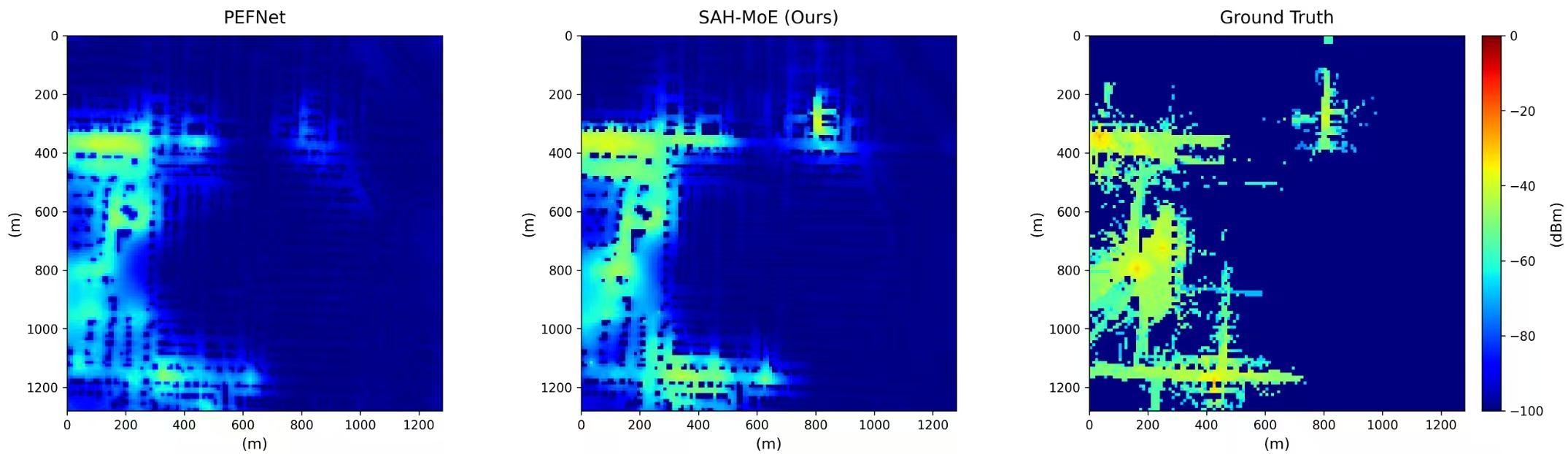}}
  \hfill
  \subfigure[{\scriptsize Dense Urban (22 GHz)}]{\includegraphics[width=0.3\textwidth]{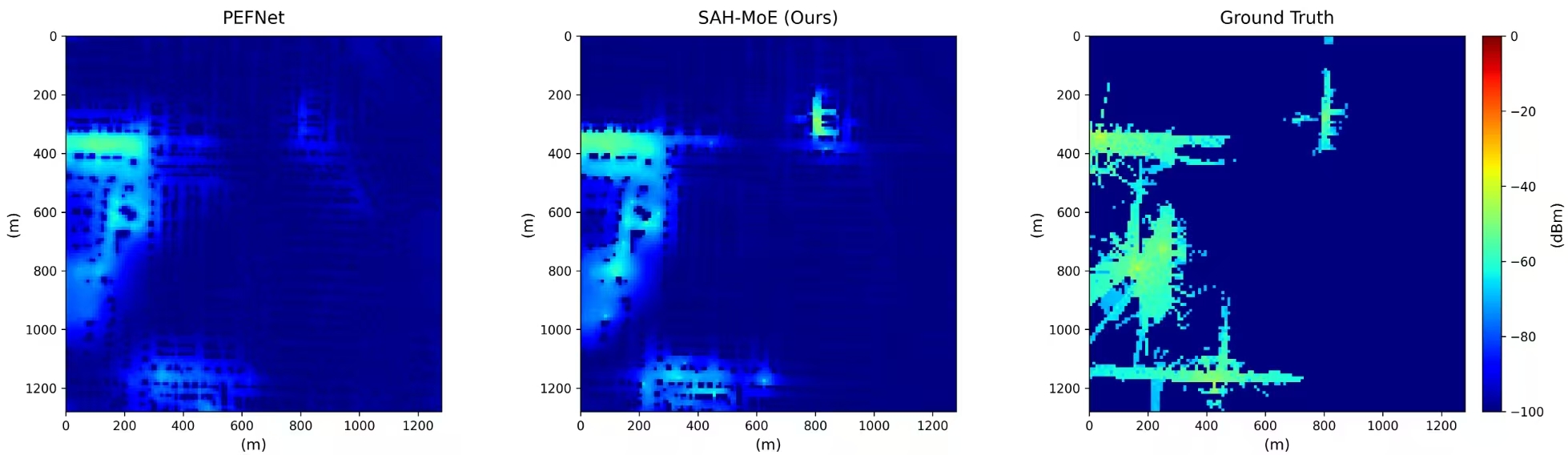}}
  \vspace{-2mm}
  \caption{Visualization of reconstructed radio maps across diverse scenarios (Ocean, Island, Dense Urban) and frequency bands (150 MHz, 1.7 GHz, 22 GHz). In each subfigure, from left to right: PEFNet, LLM-MoE (Ours), and Ground Truth.}
  \label{fig:visual_comparison_matrix}
  \vspace{-2mm}
\end{figure}

\begin{table*}[tb]
  \centering
   \vspace{-5mm}
  \caption{Performance Evaluation across Different Frequency Bands} 
  \label{tab:freq_eval_styled}
  \renewcommand{\arraystretch}{1.2}
  \resizebox{0.92\linewidth}{!}{
  \begin{tabular}{lcccccccccc} 
    \toprule
    \textbf{Methods} & \multicolumn{2}{c}{\textbf{150 MHz}} & \multicolumn{2}{c}{\textbf{1.5 GHz}} & \multicolumn{2}{c}{\textbf{1.7 GHz}} & \multicolumn{2}{c}{\textbf{3.5 GHz}} & \multicolumn{2}{c}{\textbf{22 GHz}} \\
    \cmidrule(lr){2-3} \cmidrule(lr){4-5} \cmidrule(lr){6-7} \cmidrule(lr){8-9} \cmidrule(lr){10-11}
    & NMSE$\downarrow$ & PSNR$\uparrow$ & NMSE$\downarrow$ & PSNR$\uparrow$ & NMSE$\downarrow$ & PSNR$\uparrow$ & NMSE$\downarrow$ & PSNR$\uparrow$ & NMSE$\downarrow$ & PSNR$\uparrow$ \\
    \midrule
    AE        & 0.2982 & 15.57 & 0.3040 & 17.45 & 0.3045 & 17.55 & 0.3077 & 18.16 & 0.3515 & 20.02 \\
    RadioUNet & 0.2635 & 15.96 & 0.2642 & 17.88 & 0.2645 & 17.98 & 0.2663 & 18.57 & 0.3020 & 20.45 \\
    PMNet     & 0.2729 & 15.44 & 0.2775 & 17.61 & 0.2780 & 17.72 & 0.2814 & 18.39 & 0.3240 & 20.26 \\
    PEFNet    & \underline{0.2620} & 15.97 & \underline{0.2618} & \underline{17.91} & \underline{0.2619} & \underline{18.01} & \underline{0.2634} & \underline{18.61} & \underline{0.2976} & \underline{20.48} \\
    PhyRMDM   & 0.2652 & \underline{15.99} & 0.2658 & 17.90 & 0.2660 & 18.00 & 0.2674 & 18.59 & 0.3016 & 20.47 \\
    \textbf{LLM-MoE (Ours)} & \textbf{0.2346} & \textbf{17.69} & \textbf{0.2375} & \textbf{19.62} & \textbf{0.2383} & \textbf{19.68} & \textbf{0.2403} & \textbf{20.19} & \textbf{0.2650} & \textbf{21.41} \\
    \midrule
    \textbf{Improv.} & \textbf{10.5\%} & \textbf{10.6\%} & \textbf{9.3\%} & \textbf{9.5\%} & \textbf{9.0\%} & \textbf{9.3\%} & \textbf{8.8\%} & \textbf{8.5\%} & \textbf{11.0\%} & \textbf{4.5\%} \\
    \bottomrule
  \end{tabular}
  }
  \vspace{-5mm}
\end{table*}

\begin{table*}[tb]
  \centering
  \vspace{-2mm}
  \caption{Performance Evaluation across Diverse Scenarios}
  \label{tab:scene_eval_final}
  % 第一部分：前 5 个场景
  \resizebox{0.92\textwidth}{!}{
  \begin{tabular}{lcccccccccc} 
    \toprule
    \textbf{Methods} & \multicolumn{2}{c}{\textbf{Grassland}} & \multicolumn{2}{c}{\textbf{Island}} & \multicolumn{2}{c}{\textbf{Ocean}} & \multicolumn{2}{c}{\textbf{Lake}} & \multicolumn{2}{c}{\textbf{Ordinary Urban}} \\
    \cmidrule(lr){2-3} \cmidrule(lr){4-5} \cmidrule(lr){6-7} \cmidrule(lr){8-9} \cmidrule(lr){10-11}
    & NMSE$\downarrow$ & PSNR$\uparrow$ & NMSE$\downarrow$ & PSNR$\uparrow$ & NMSE$\downarrow$ & PSNR$\uparrow$ & NMSE$\downarrow$ & PSNR$\uparrow$ & NMSE$\downarrow$ & PSNR$\uparrow$ \\
    \midrule
    AE        & 0.2177 & 16.03 & 0.1923 & 17.61 & 0.0404 & 21.35 & 0.0976 & 21.24 & 0.3488 & 17.25 \\
    RadioUNet & 0.2130 & 16.14 & 0.1870 & 17.81 & 0.0360 & 21.92 & 0.0951 & 21.29 & 0.2803 & 18.18 \\
    PMNet     & \underline{0.2089} & \underline{16.20} & 0.1934 & 17.52 & 0.0393 & 21.38 & 0.1004 & 20.86 & 0.3159 & 17.66 \\
    PEFNet    & 0.2130 & 16.13 & 0.1872 & 17.80 & \underline{0.0358} & \underline{21.95} & 0.0947 & 21.31 & \underline{0.2770} & \underline{18.21} \\
    PhyRMDM   & 0.2145 & 16.13 & \underline{0.1869} & \underline{17.84} & 0.0360 & \underline{21.95} & \underline{0.0938} & \underline{21.39} & 0.2826 & \underline{18.21} \\
    \textbf{LLM-MoE (Ours)} & \textbf{0.2018} & \textbf{16.37} & \textbf{0.1807} & \textbf{18.38} & \textbf{0.0069} & \textbf{32.63} & \textbf{0.0856} & \textbf{24.79} & \textbf{0.2564} & \textbf{18.89} \\
    \midrule
    \textbf{Improv.} & \textbf{3.4\%} & \textbf{1.0\%} & \textbf{3.3\%} & \textbf{3.0\%} & \textbf{80.7\%} & \textbf{48.7\%} & \textbf{8.7\%} & \textbf{15.9\%} & \textbf{7.4\%} & \textbf{3.7\%} \\
    \bottomrule
  \end{tabular}
  }  
  % 第二部分：后 5 个场景
  \resizebox{0.92\textwidth}{!}{
  \begin{tabular}{lcccccccccc} 
    \toprule
    \textbf{Methods} & \multicolumn{2}{c}{\textbf{Rural}} & \multicolumn{2}{c}{\textbf{Desert}} & \multicolumn{2}{c}{\textbf{Mountainous}} & \multicolumn{2}{c}{\textbf{Suburban}} & \multicolumn{2}{c}{\textbf{Dense Urban}} \\
    \cmidrule(lr){2-3} \cmidrule(lr){4-5} \cmidrule(lr){6-7} \cmidrule(lr){8-9} \cmidrule(lr){10-11}
    & NMSE$\downarrow$ & PSNR$\uparrow$ & NMSE$\downarrow$ & PSNR$\uparrow$ & NMSE$\downarrow$ & PSNR$\uparrow$ & NMSE$\downarrow$ & PSNR$\uparrow$ & NMSE$\downarrow$ & PSNR$\uparrow$ \\
    \midrule
    AE        & 0.2343 & 15.27 & 0.3661 & 15.53 & 0.2377 & 16.84 & 0.2071 & 16.19 & 0.5649 & 17.27 \\
    RadioUNet & \underline{0.2083} & \underline{15.76} & 0.3650 & 15.57 & 0.2342 & 16.90 & 0.1994 & 16.37 & 0.4308 & 18.31 \\
    PMNet     & 0.2323 & 15.30 & \underline{0.3483} & \underline{15.71} & \underline{0.2285} & \underline{16.96} & 0.2041 & 16.23 & 0.4842 & 17.80 \\
    PEFNet    & 0.2084 & \underline{15.76} & 0.3646 & 15.57 & 0.2348 & 16.90 & \underline{0.1988} & \underline{16.38} & \underline{0.4214} & \underline{18.39} \\
    PhyRMDM   & 0.2096 & 15.73 & 0.3677 & 15.55 & 0.2379 & 16.85 & 0.1998 & 16.37 & 0.4326 & 18.30 \\
    \textbf{LLM-MoE (Ours)} & \textbf{0.1993} & \textbf{15.94} & \textbf{0.3233} & \textbf{15.96} & \textbf{0.2204} & \textbf{17.10} & \textbf{0.1924} & \textbf{16.57} & \textbf{0.3775} & \textbf{18.86} \\
    \midrule
    \textbf{Improv.} & \textbf{4.3\%} & \textbf{1.1\%} & \textbf{7.2\%} & \textbf{1.6\%} & \textbf{3.5\%} & \textbf{0.8\%} & \textbf{3.2\%} & \textbf{1.2\%} & \textbf{10.4\%} & \textbf{2.6\%} \\
    \bottomrule
  \end{tabular}
  }
  \vspace{-5mm}
\end{table*}

\begin{table}[tb]
  \centering
  \caption{Comprehensive Ablation Study on Core Components}
  \vspace{-2mm}
  \scriptsize
  \label{tab:ablation_study}
  \begin{tabular*}{0.92\linewidth}{@{\extracolsep{\fill}}lcccc} 
    \toprule
    \textbf{Model / Configuration} & \textbf{MSE$\downarrow$} & \textbf{RMSE$\downarrow$} & \textbf{NMSE$\downarrow$} & \textbf{PSNR$\uparrow$} \\
    \midrule
    \multicolumn{5}{c}{\textit{Modality Ablation}} \\
    \midrule
    LLM-MoE w/o Visual Features    & 0.01836 & 0.12290 & 0.24640 & 19.58 \\
    LLM-MoE w/o Semantic Guidance  & 0.01921 & 0.12827 & 0.25398 & 18.70 \\
    \midrule
    \multicolumn{5}{c}{\textit{Single Expert Ablation (Leave-One-Out)}} \\
    \midrule
    LLM-MoE w/o CNN Expert         & 0.01977 & 0.12879 & 0.26320 & 18.90 \\
    LLM-MoE w/o ViT Expert         & 0.01828 & 0.12282 & 0.24379 & 19.50 \\
    LLM-MoE w/o MLP Expert         & 0.01848 & 0.12376 & 0.24662 & 19.43 \\
    LLM-MoE w/o GNN Expert         & 0.01827 & 0.12305 & 0.24396 & 19.45 \\
    \midrule
    \multicolumn{5}{c}{\textit{Dual Expert Ablation (Leave-Two-Out)}} \\
    \midrule
    LLM-MoE w/o CNN \& ViT         & 0.02003 & 0.12904 & 0.26882 & 18.95 \\
    LLM-MoE w/o CNN \& MLP         & 0.01963 & 0.12713 & 0.26599 & 19.26 \\
    LLM-MoE w/o CNN \& GNN         & 0.01900 & 0.12623 & 0.25196 & 19.12 \\
    LLM-MoE w/o ViT \& MLP         & 0.01845 & 0.12348 & 0.24630 & 19.49 \\
    LLM-MoE w/o ViT \& GNN         & 0.01825 & 0.12249 & 0.24387 & 19.62 \\
    LLM-MoE w/o MLP \& GNN         & 0.01841 & 0.12304 & 0.24667 & 19.57 \\
    \midrule
    \textbf{Full LLM-MoE (Ours)}   & \textbf{0.01819} & \textbf{0.12183} & \textbf{0.24313} & \textbf{19.72} \\
    \bottomrule
  \end{tabular*}
  \vspace{-5mm}
\end{table}

\textbf{Performance Across Frequencies.} We evaluate the performance across different frequencies. As shown in Table~\ref{tab:freq_eval_styled}, LLM-MoE consistently achieves the best performance across all frequency bands. It achieves the lowest NMSE and the highest PSNR from 150 MHz to 22 GHz, demonstrating strong cross-frequency performance. Compared with PEFNet, LLM-MoE achieves consistent improvements across all bands, with NMSE reductions ranging from 8.8\% to 11.0\%. These stable gains indicate that the model effectively captures frequency-dependent propagation characteristics. We attribute this to the LLM-guided gating mechanism that adaptively assigns frequency-specific modeling tasks to different experts. Fig.~{\ref{fig:visual_comparison_matrix}} provides qualitative comparisons across frequencies. LLM-MoE produces radio maps that are consistently closer to the ground truth. At low frequency, such as 150 MHz, it better captures propagation patterns with fewer artifacts, as shown in Fig.~{\ref{fig:visual_comparison_matrix}} (a). At higher frequencies, such as 22 GHz, it more accurately reconstructs sharp attenuation boundaries, as shown in Fig.~{\ref{fig:visual_comparison_matrix}} (i). In contrast, PEFNet exhibits artifacts at low frequencies and fails to preserve fine-grained structures at higher frequencies, further confirming LLM-MoE's superior adaptability across frequencies.

\textbf{Performance Across Scenarios.} We evaluate the performance of radio map construction across various scenarios. Table~\ref{tab:scene_eval_final} summarizes the results. LLM-MoE consistently achieves the best performance in all scenarios. Notably, even in complex environments such as Dense Urban, it attains the lowest NMSE and highest PSNR, outperforming the second-best baseline by 10.4\% and 2.6\%, respectively. These results demonstrate the strong adaptability of LLM-MoE across diverse scenarios. Fig.~\ref{fig:visual_comparison_matrix} further provides qualitative comparisons. LLM-MoE generates radio maps that are consistently closer to the ground truth. In scenarios dominated by line-of-sight (LoS) propagation, such as Ocean, it accurately preserves smooth large-scale propagation patterns. In contrast, in Dense Urban environments with strong blockage and attenuation, it more effectively reconstructs sharp attenuation regions and intricate structural variations. The baseline method exhibits noticeable artifacts in LoS-dominated scenarios and fails to capture fine-grained structures under severe blockage conditions. These observations further confirm the superior capability of LLM-MoE in modeling diverse propagation environments.

\subsection{Ablation Study}
We conduct ablation studies to evaluate the contributions of the core components, as summarized in Table~\ref{tab:ablation_study}. In the modality ablation, removing either semantic guidance or visual features leads to clear performance degradation. This demonstrates that accurate expert weight allocation relies on the joint regulation of both macroscopic semantics and local visual observation. For the expert ablation, removing any single expert consistently degrades performance. For example, excluding the CNN expert increases the MSE to 0.01977, while removing both CNN and ViT experts further degrades the performance to an MSE of 0.02003. Similar trends can be observed for other expert combinations. Overall, the performance deterioration under both single and dual expert removal verifies that each expert contributes complementary knowledge, and the full heterogeneous expert pool is essential for achieving optimal performance.
\section{Conclusion}
\label{sec:conclusion}

In this paper, we propose an LLM-MoE framework for radio map prediction in complex multi-band and multi-scenario environments. The heterogeneous expert pool, comprising CNN, ViT, MLP, and GNN, effectively fits distinct physical propagation mechanisms. We introduced an LLM to extract global semantic descriptions and designed a Cross-Modal Gating Network, which leverages cross-attention to link macroscopic semantics with local spatial features, thereby enabling adaptive expert weight allocation. Extensive experiments on the SpectrumNet dataset demonstrate that the LLM-MoE framework significantly reduces prediction errors, maintaining robust generalization across diverse scenarios and frequency bands.

\section*{Acknowledgments}
This research was supported in part by the National Natural Science Foundation of China under Grant U24A20247 and Grant 62471277, the Young Elite Scientists Sponsorship Program by CAST under Grant ZB2025-293, the Fundamental Research Funds for the Central Universities under Grant 531119200354, and the Future Industry Technology Innovation Special Project of the Key Science and Technology Support Program of Hebei Province under Grant 2050201.

\bibliographystyle{splncs04}
\bibliography{mybibliography} % 确保上传并链接 mybibliography.bib 

\end{document}